\documentclass[aps,prb,twocolumn,showpacs]{revtex4}

\usepackage{graphicx}

\begin{document}

\title{Resonance peak in underdoped cuprates}

\author{A.~Sherman}

\affiliation{Institute of Physics, University of Tartu, Riia 142, 51014
Tartu, Estonia}

\author{M.~Schreiber}

\affiliation{Institut f\"ur Physik, Technische Universit\"at, D-09107
Chemnitz, Federal Republic of Germany}

\date{\today}

\begin{abstract}
The magnetic susceptibility measured in neutron scattering experiments
in underdoped YBa$_2$Cu$_3$O$_{7-y}$ is interpreted based on the
self-consistent solution of the $t$-$J$ model of a Cu-O plane. The
calculations reproduce correctly the frequency and momentum
dependencies of the susceptibility and its variation with doping and
temperature in the normal and superconducting states. This allows us to
interpret the maximum in the frequency dependence -- the resonance peak
-- as a manifestation of the excitation branch of localized Cu spins
and to relate the frequency of the maximum to the size of the spin gap.
The low-frequency shoulder well resolved in the susceptibility of
superconducting crystals is connected with a pronounced maximum in the
damping of the spin excitations. This maximum is caused by intense
quasiparticle peaks in the hole spectral function for momenta near the
Fermi surface and by the nesting.
\end{abstract}


\maketitle

\section{Introduction}
Inelastic neutron scattering experiments give important information on
the anomalous properties of high-$T_c$ superconductors. Among the
results obtained with this experimental method is the detailed
information on the magnetic susceptibility in YBa$_2$Cu$_3$O$_{7-y}$
measured in wide ranges of hole concentrations and temperatures.
\cite{Rossat,Bourges} These measurements revealed the sharp magnetic
collective mode called the resonance peak. The peak first observed
\cite{Regnault,Mook} in the superconducting state of YBa$_2$Cu$_3$O$_7$
was later also detected in the underdoped compounds, both in the
superconducting and normal states. \cite{Bourges,Dai} Recently the
resonance peak was also observed in Bi$_2$Sr$_2$CaCu$_2$O$_{8+\delta}$
and Tl$_2$Ba$_2$CuO$_{6+\delta}$. \cite{He}

Theoretical works devoted to the resonance peak were mainly
concentrated at the overdoped region where the peak is observed in the
superconducting state and disappears in the normal state.
\cite{Rossat,Bourges,Regnault,Mook,Dai,He} In
Refs.~\onlinecite{Liu,Mazin,Bulut} an interpretation of the resonance
peak based on the itinerant magnetism approach was proposed. This
approach which uses the Lindhard function for the bare susceptibility
$\chi_0({\bf k}\omega)$ and the random phase approximation relates the
appearance of the peak to the disappearance or considerable decrease of
${\rm Im}\chi_0({\bf Q}\omega)$ in the frequency range $\omega \leq
2\Delta^s$ with the opening of the $d$-wave \cite{Liu,Bulut} or
$s$-wave \cite{Mazin} superconducting gaps $\Delta^s$. Here ${\bf
Q}=(\pi,\pi)$ is the antiferromagnetic wave vector. In this frequency
range the peak arises due to the logarithmic divergence in ${\rm
Re}\chi_0({\bf Q}\omega)$ which originates from the jump
\cite{Liu,Mazin} in ${\rm Im}\chi_0({\bf Q}\omega)$ or due to the
nesting of the bonding and antibonding Fermi surfaces in the two-layer
crystal. \cite{Bulut} In the normal state, when the damping increases,
the peak is smeared out. For the approaches of
Refs.~\onlinecite{Mazin,Bulut} the two-layer structure of
YBa$_2$Cu$_3$O$_{7-y}$ is of crucial importance for the appearance of
the resonance peak. However, the recent observation \cite{He} of the
peak in single-layer Tl$_2$Ba$_2$CuO$_{6+\delta}$ indicates that an
interaction between closely spaced Cu-O layers is not the necessary
condition. A single-layer system described by a modified $s$-$f$
Hamiltonian was considered in Ref.~\onlinecite{Abanov}. In that work
the resonance peak appears also due to the vanishing damping of spin
excitations and a real part of the spin excitation frequency stems from
the fermion bubble. A qualitatively different approach was suggested in
Ref.~\onlinecite{Morr} where the existence of a well-defined branch of
spin excitations which exists even in the absence of mobile carriers
was postulated near the $M$ point (${\bf k=Q}$) of the Brillouin zone.
In this scenario the resonance peak is related to the excitation with
${\bf k=Q}$ of this branch. As in Ref.~\onlinecite{Liu,Mazin,Bulut},
here the peak is visible in the superconducting state due to the
absence of the damping for $\omega \leq 2\Delta^s$ and is smeared out
in the normal state.

This latter scenario seems to correspond most adequately to available
experimental data on the resonance peak. As mentioned, it is observed
also in underdoped YBa$_2$Cu$_3$O$_{7-y}$ and in the superconducting
state the peak varies continuously on passing from the underdoped to
overdoped region. Moreover, in the underdoped region the peak is also
observed in the normal state and its frequency is nearly the same as in
the superconducting state. Therefore it is reasonable to search for a
unified explanation for the peak which is applicable both for
underdoped and overdoped regions, and -- in the former region -- for
the normal and superconducting states. The existence of the excitation
branch of localized Cu spins is well established for the underdoped
region. \cite{Bourges,Birgeneau} In this region it is quite reasonable
to connect the resonance peak with these excitations.

In this paper we use the two-dimensional $t$-$J$ model to which the
realistic three-band Hubbard model of the CuO$_2$ planes can be mapped
in the case of a strong on-site Coulomb repulsion. \cite{Jefferson} For
the underdoped region the self-consistent solution of the $t$-$J$ model
was obtained in Ref.~\onlinecite{Sherman02} with the use of Mori's
projection operator technique. We employ this result for the
calculation of the magnetic susceptibility in the normal and
superconducting states. The calculations reproduce correctly the
frequency and momentum dependencies of the susceptibility and its
evolution with doping and temperature in YBa$_2$Cu$_3$O$_{7-y}$. This
allows us to relate the resonance peak with the excitation branch of
the localized Cu spins and to identify the frequency of the peak with
the size of the spin gap at the $M$ point. In the underdoped region,
with increasing doping the peak frequency grows with the gap size and
the peak intensity decreases, in agreement with experimental
observations. \cite{Rossat,Bourges} Moreover, the low-frequency
shoulder observed \cite{Rossat,Bourges} in the susceptibility of
superconducting crystals can be connected with a pronounced maximum
which we find in the damping of the spin excitations. This maximum is
caused by intense quasiparticle peaks in the hole spectral function for
momenta near the Fermi surface and by the nesting.

\section{The hole Green's function in the superconducting state}
The Hamiltonian of the 2D $t$-$J$ model reads \cite{Izyumov}
\begin{equation}
H=\sum_{\bf nm\sigma}t_{\bf nm}a^\dagger_{\bf n\sigma}a_{\bf
m\sigma}+\frac{1}{2}\sum_{\bf nm}J_{\bf nm}\left(s^z_{\bf n}s^z_{\bf
m}+s^{+1}_{\bf n}s^{-1}_{\bf m}\right), \label{hamiltonian}
\end{equation}
where $a_{\bf n\sigma}=|{\bf n}\sigma\rangle\langle{\bf n}0|$ is the
hole annihilation operator, {\bf n} and {\bf m} label sites of the
square lattice, $\sigma=\pm 1$ is the spin projection, $|{\bf
n}\sigma\rangle$ and $|{\bf n}0\rangle$ are site states corresponding
to the absence and presence of a hole on the site. These states are
linear combinations of the products of the $3d_{x^2-y^2}$ copper and
$2p_\sigma$ oxygen orbitals of the extended Hubbard model.
\cite{Jefferson} In this work we take into account nearest neighbor
interactions only, $t_{\bf nm}=-t\sum_{\bf a}\delta_{\bf n,m+a}$ and
$J_{\bf nm}=J\sum_{\bf a}\delta_{\bf n,m+a}$ where the four vectors
{\bf a} connect nearest neighbor sites. The spin-$\frac{1}{2}$
operators can be written as $s^z_{\bf
n}=\frac{1}{2}\sum_\sigma\sigma|{\bf n}\sigma\rangle\langle{\bf
n}\sigma|$ and $s^\sigma_{\bf n}=|{\bf n}\sigma\rangle\langle{\bf
n},-\sigma|$.

To investigate the magnetic susceptibility of this model the hole ${\bf
G}({\bf k}t)=-i\theta(t)\langle\!\{{\bf A}_{\bf k\sigma}(t),{\bf
A}^\dagger_{\bf k\sigma}\}\!\rangle$ and spin $D({\bf
k}t)=-i\theta(t)\langle[s^z_{\bf k}(t), s^z_{\bf -k}]\rangle$ Green's
functions have to be calculated. Here supposing the singlet
superconducting pairing we introduced the Nambu spinor,
$${\bf A_{\bf k\sigma}}=\pmatrix{a_{\bf k\sigma}\cr a^\dagger_{\bf
-k,-\sigma}},$$ thus ${\bf G}$ is a 2$\times$2 matrix (here and below
matrices and vectors are designated by boldface letters). In the above
formulas the angular brackets denote averaging over the grand canonical
ensemble, and
\begin{eqnarray*}
a_{\bf k\sigma}&=&N^{-1/2}\sum_{\bf n}e^{-i{\bf kn}}a_{\bf
n\sigma},\\
s^z_{\bf k}&=&N^{-1/2}\sum_{\bf n}e^{-i{\bf kn}}s^z_{\bf n},
\end{eqnarray*}
$a_{\bf k\sigma}(t)=\exp(i{\cal H}t)a_{\bf k\sigma}\exp(-i{\cal H}t)$,
$N$ is the number of sites, ${\cal H}=H-\mu\sum_{\bf n}X_{\bf n}$,
$\mu$ is the chemical potential, $X_{\bf n}=|{\bf n}0\rangle
\langle{\bf n}0|$.

To derive the self-energy equation for the matrix Green's function
${\bf G}$ the continued fraction representation for Green's function
and the recursive equations for their elements from
Ref.~\onlinecite{Sherman02} have to be generalized for the case of
matrices. Such generalization reads
\begin{equation}
{\bf R}_n(\omega)=[\omega{\bf I}-{\bf E}_n-{\bf R}_{n+1}(\omega){\bf
F}_n]^{-1},\quad n=0,1,2\ldots \label{Mori}
\end{equation}
where ${\bf I}$ is a 2$\times$2 unit matrix, the matrices ${\bf E}_n$
and ${\bf F}_n$ are calculated from the recursive equations
\begin{eqnarray}
&&[{\bf A}_n,H]={\bf E}_n{\bf A}_n+{\bf A}_{n+1}+{\bf F}_{n-1}{\bf
A}_{n-1}, \nonumber\\
&&{\bf E}_n=\langle\{[{\bf A}_n,H],{\bf A}^\dagger_n\}\rangle
\langle\{{\bf A}_n,{\bf A}^\dagger_n\}\rangle^{-1},\label{Lanczos}\\
&&{\bf F}_n=\langle\{{\bf A}_{n+1},{\bf A}^\dagger_{n+1}\}\rangle
\langle\{{\bf A}_n,{\bf A}^\dagger_n\}\rangle^{-1}. \nonumber
\end{eqnarray}
Here ${\bf F}_{-1}=0$ and ${\bf A}_0={\bf A}_{\bf k\sigma}$ for the
case of the function ${\bf G}$. As follows from Eq.~(\ref{Lanczos}),
the two-component operators ${\bf A}_n$ constructed in this recursive
procedure form an orthogonal basis. For the anticommutator Green's
function ${\bf G}$ the inner product of two arbitrary operators ${\bf
A}$ and ${\bf B}$ is defined as $\langle\{{\bf A},{\bf B}\}\rangle$ and
the orthogonality means $\langle\{{\bf A}_n,{\bf
A}_m\}\rangle=\delta_{nm}\langle\{{\bf A}_n,{\bf A}_n\}\rangle$. In
Eq.~(\ref{Mori}), ${\bf R}_n(\omega)=-i\int^\infty_0 dt \exp(i\omega
t){\bf R}_n(t)$, ${\bf R}_n(t)=\langle\{{\bf A}_{nt},{\bf
A}^\dagger_{n}\}\rangle\langle\{{\bf A}_{n},{\bf
A}^\dagger_{n}\}\rangle^{-1}$ where the time dependencies are
determined by the equation
$$i\frac{d}{dt}{\bf A}_{nt}=\prod^{n-1}_{k=0}(1-P_k)[{\bf A}_{nt},H],
\quad {\bf A}_{n,t=0}={\bf A}_n$$ with the definition $P_n{\bf
Q}=\langle\{{\bf Q},{\bf A}^\dagger_n\}\rangle \langle\{{\bf A}_n,{\bf
A}^\dagger_n\}\rangle^{-1}{\bf A}_n$ of the projection operator $P_n$
that projects an arbitrary two-component operator ${\bf Q}$ on ${\bf
A}_n$ (for a more detailed discussion of these equations see
Ref.~\onlinecite{Sherman02}).

From the above definitions it follows for the Fourier transformation of
${\bf G}({\bf k}t)$
\begin{equation}
{\bf G}({\bf k}\omega)=[\omega{\bf I}-{\bf E}_0-{\bf R}_1{\bf
F}_0]^{-1}\langle\{{\bf A}_0,{\bf A}^\dagger_0\}\rangle,\label{hgf}
\end{equation}
where $\langle\{{\bf A}_0,{\bf A}^\dagger_0\}\rangle=\varphi{\bf I}$,
$\varphi=\frac{1}{2}(1+x)$, $x=\langle X_{\bf n}\rangle$ is the hole
concentration,
\begin{equation}
{\bf E}_0=\pmatrix{\varepsilon_{\bf k}-\mu' & \sigma K_1\varphi^{-1}
(3J\gamma_{\bf k}-8t) \cr \sigma K_1^*\varphi^{-1} (3J\gamma_{\bf
k}-8t) & -(\varepsilon_{\bf k}-\mu')}, \label{e0}
\end{equation}
$\varepsilon_{\bf
k}=-(4t\varphi+6tC_1\varphi^{-1}+3JF_1\varphi^{-1})\gamma_{\bf k}$,
$\mu'=\mu+4tF_1\varphi^{-1}+3JC_1\varphi^{-1}$, $\gamma_{\bf
k}=\frac{1}{4}\sum_{\bf a}\exp(i{\bf ka})$. The nearest-neighbor
correlations $C_1=\langle s^{+1}_{\bf n}s^{-1}_{\bf n+a}\rangle$,
$F_1=\langle a^\dagger_{\bf n\sigma}a_{\bf n+a,\sigma}\rangle$,
$K_1=\sigma\langle a_{\bf n\sigma}a_{\bf n+a,-\sigma}\rangle$ and the
hole concentration $x$ can be expressed in terms of the components of
the hole and spin Green's functions:
\begin{eqnarray}
x&=&\frac{1}{N}\sum_{\bf k}\int_{-\infty}^\infty d\omega
 n_F(\omega)A({\bf k}\omega),\nonumber\\
F_1&=&\frac{1}{N}\sum_{\bf k}\gamma_{\bf k}\int_{-\infty}^\infty
 d\omega n_F(\omega)A({\bf k}\omega),\nonumber\\
C_1&=&\frac{2}{N}\sum_{\bf k}\gamma_{\bf k}\int_0^\infty d\omega
 \coth\!\left(\frac{\omega}{2T}\right)B({\bf k}\omega),
 \label{xcf}\\
K_1&=&\frac{\sigma}{N}\sum_{\bf k}\gamma_{\bf k}\int_{-\infty}^\infty
 d\omega [1-n_F(\omega)]\nonumber\\
 &&\times[L({\bf k}\omega\sigma)+iM({\bf k}\omega\sigma)],\nonumber
\end{eqnarray}
where
\begin{eqnarray*}
L({\bf k}\omega\sigma)&=&-{\rm Im}\,[G_{12}({\bf k}\omega\sigma)
+G_{21}({\bf k}\omega\sigma)]/(2\pi),\\
M({\bf k}\omega\sigma)&=&{\rm Re}\,[G_{12}({\bf k}\omega\sigma)
-G_{21}({\bf k}\omega\sigma)]/(2\pi),
\end{eqnarray*}
$A({\bf k}\omega)=-{\rm Im}\,G_{11}({\bf k}\omega)/\pi$, and $B({\bf
k}\omega)=-{\rm Im}\,D({\bf k}\omega)/\pi$ are the hole and spin
spectral functions, $n_F(\omega)=[\exp(\omega/T)+1]^{-1}$ and $T$ is
the temperature (the functions $A({\bf k}\omega)$ and $B({\bf
k}\omega)$ do not depend on $\sigma$). In the derivation of
Eq.~(\ref{xcf}) for $C_1$ we have taken into account that the
approximation used retains the rotation symmetry of spin components
\cite{Sherman02} and therefore $C_1=2\langle s^z_{\bf n}s^z_{\bf
n+a}\rangle$.

As follows from Eq.~(\ref{e0}), in the $t$-$J$ model the
superconducting gap has an $s$-wave component if $K_1 \neq 0$. However,
in the considered case this component is small in comparison with the
$d$-wave component introduced below and will be neglected.

From the definition of the hole self-energy ${\bf\Sigma=R_1F_0}$ we
find that $\Sigma_{22}({\bf k}\omega)=\Sigma^*_{11}({\bf k},-\omega)$
where it was taken into account that these components of ${\bf\Sigma}$
do not depend on $\sigma$ and are invariant under the inversion of
${\bf k}$. For $\Sigma_{11}({\bf k}\omega)$ the following expression
obtained in Ref.~\onlinecite{Sherman02} can be used:
\begin{eqnarray}
{\rm Im}\,\Sigma_{11}({\bf k}\omega)&=&\frac{16\pi t^2}{N\phi}\sum_{\bf
 k'}\int_{-\infty}^\infty d\omega'\biggl[\gamma_{\bf k-k'}+\gamma_{\bf
 k} \nonumber\\
&+&{\rm sgn}(\omega')(\gamma_{\bf
 k-k'}-\gamma_{\bf k})\sqrt{\frac{1+\gamma_{\bf k'}}{1-\gamma_{\bf
 k'}}}\biggr]^2 \nonumber\\
&\times&[n_B(-\omega')+n_F(\omega-\omega')] \label{se}\\
&\times&A({\bf k-k'},\omega-\omega')B({\bf k'}\omega'), \nonumber\\
{\rm Re}\,\Sigma_{11}({\bf k}\omega)&=&{\cal P}\int^\infty_{-\infty}
 \frac{d\omega'}{\pi}\frac{{\rm Im}\,\Sigma_{11}({\bf
 k}\omega')}{\omega'-\omega}\nonumber,
\end{eqnarray}
where $n_B(\omega)=\left[ \exp(\omega/T)-1\right]^{-1}$ and ${\cal P}$
indicates Cau\-chy's principal value.

Assuming the $d$-wave superconducting pairing, for the anomalous
self-energies we set
\begin{equation}
\Sigma_{12}({\bf k}\omega\sigma)=\Sigma_{12}({\bf
k}\omega\sigma)=\sigma\Delta^s[\cos(k_x)-\cos(k_y)]/2,\label{ase}
\end{equation}
with the superconducting gap $\Delta^s$. For such anomalous
self-energies $M({\bf k}\omega\sigma)=0$.

\section{The spin Green's function}
In Ref.~\onlinecite{Sherman02} we have noticed that the approximation
used there leads to an underestimation of the imaginary part of the
magnetic susceptibility at low frequencies. To avoid this drawback in
the present work we shall not split the spin self-energy into the hole
and spin parts, but rather continue the calculation of the terms of the
continued fraction using the entire Hamiltonian (\ref{hamiltonian}).

The spin Green's function is calculated from the relation
\begin{equation}
D({\bf k}\omega)=\omega((s^z_{\bf k}|s^z_{\bf -k}))_\omega-(s^z_{\bf
k},s^z_{\bf -k}), \label{gfmr}
\end{equation}
where
\begin{equation}
(s^z_{\bf k},s^z_{\bf -k})=i\int^\infty_0dt\langle[s^z_{\bf
k}(t),s^z_{\bf -k}]\rangle, \label{inprod}
\end{equation}
and Kubo's relaxation function
\begin{equation}
((s^z_{\bf k}|s^z_{\bf -k}))_\omega=\int^\infty_0dte^{i\omega
t}\int^\infty_tdt'\langle[s^z_{\bf k}(t'),s^z_{\bf -k}]\rangle
\label{Kubo}
\end{equation}
can be represented by a continued fraction which is similar to the
scalar form of Eq.~(\ref{Mori}). The elements $E_n$ and $F_n$ of this
function are calculated from a recursive procedure which is similar to
the scalar form of Eq.~(\ref{Lanczos}) where, however, mean values of
anticommutators have to be substituted by inner products of the type of
Eq.~(\ref{inprod}) (see Ref.~\onlinecite{Sherman02}).

From this definition we find for the starting operator $A_0=s^z_{\bf
k}$ of this recursive procedure
$$E_0=(i\dot{s}^z_{\bf k},s^z_{\bf -k})(s^z_{\bf k},s^z_{\bf
-k})^{-1}=0,$$ where $i\dot{s}^z_{\bf k}=[s^z_{\bf k},H]$,
$$A_1=i\dot{s}^z_{\bf k}, \quad F_0=\frac{4(1-\gamma_{\bf
k})(J|C_1|+tF_1)}{(s^z_{\bf k},s^z_{\bf -k})}, \quad E_1=0.$$ Using
these elements of the continued fraction representation of $((s^z_{\bf
k}|s^z_{\bf -k}))_\omega$, Eq.~(\ref{gfmr}) can be rewritten as
\begin{equation}
D({\bf k}\omega)=\frac{4(1-\gamma_{\bf
k})(J|C_1|+tF_1)}{\omega^2-\omega\Pi({\bf k}\omega)-\omega^2_{\bf k}},
\label{sgf}
\end{equation}
where $\omega^2_{\bf k}=F_0$,
\begin{eqnarray*}
\Pi({\bf k}\omega)&=&-i[4(1-\gamma_{\bf k})(J|C_1|+tF_1)]^{-1}\\
&&\times\int^\infty_0dte^{i\omega t}(A_{2t},A^\dagger_2),\\
A_2&=&i^2\ddot{s}^z_{\bf k}-\omega^2_{\bf k}s^z_{\bf k}.
\end{eqnarray*}

As follows from the above equation, to calculate $\omega^2_{\bf k}$ and
$A_2$ we have to select terms of $i^2\ddot{s}^z_{\bf k}$ which are
proportional to $s^z_{\bf k}$. It can be done only approximately
because the quantity $(s^z_{\bf k},s^z_{\bf -k})$ cannot be calculated
exactly. Following Refs.~\onlinecite{Kondo,Sherman02} we used the
decoupling in $i^2\ddot{s}^z_{\bf k}$ for such selection and found
\begin{equation}
\omega^2_{\bf k}=16\alpha J^2\left(|C_1|+\frac{tF_1}{\alpha
J}\right)(1-\gamma_{\bf k})(\Delta+1+\gamma_{\bf k}), \label{frequency}
\end{equation}
where $\Delta$ is the parameter of the gap in the spin excitation
spectrum at the wave vector ${\bf Q}$ of the Brillouin zone. In an
infinite 2D lattice this gap is opened for any nonzero temperature
\cite{Kondo} and at $T=0$ for $x \agt 0.02$. \cite{Sherman02} The gap
size is directly connected with the spin correlation length of the
short-range antiferromagnetic order. Hence a finite gap for $T>0$ is in
agreement with the Mermin-Wagner theorem. \cite{Mermin} The gap
parameter can be expressed through the model parameters and
correlations of hole and spin operators. \cite{Sherman02,Kondo}
However, due to strong dependencies of the considered quantities on
this parameter we found it more accurate to determine this parameter
from the constraint of zero site magnetization $\langle s^z_{\bf
k}\rangle=0$ which is fulfilled in the paramagnetic state. This
constraint can be written in the form
\begin{equation}\label{zsm}
\frac{1}{2}(1-x)=\frac{2}{N}\sum_{\bf k}\int_0^\infty d\omega
\coth\!\left(\frac{\omega}{2T}\right)B({\bf k}\omega).
\end{equation}

In Eq.~(\ref{frequency}), the parameter $\alpha$ is introduced to
improve somewhat the results obtained with the decoupling and to take
into account vertex corrections. In earlier works \cite{Kondo} where
the analogous correction were used for the Heisenberg model this
parameter was determined from the constraint~(\ref{zsm}). Due to
comparatively weak dependencies of the considered quantities on this
parameter we found it more appropriate to set
$\alpha=1.802-0.802\tanh(10x)$ and to use the constraint for the
calculation of $\Delta$, as mentioned above. The expression given for
$\alpha$ takes into account its value obtained in
Ref.~\onlinecite{Sherman02} for finite damping of spin excitations and
the weakening of the vertex corrections with doping.

When selecting terms of $i^2\ddot{s}^z_{\bf k}$ which have to be
included into $A_2$ we omitted terms proportional to $t^2$, being
motivated by our earlier result \cite{Sherman02} and by the results of
the spin-wave approximation \cite{Sherman98} which indicate that
$\Pi({\bf k}\omega)$ has to be proportional to $t^2$. An additional
argument to omit these terms is that a part of them contains
multipliers of the type $\sum_\sigma \sigma a^\dagger_{\bf
m\sigma}a_{\bf m\sigma}$ the mean values of which are zero. Other terms
of this type and a part of terms proportional to $tJ$ contain the hole
operators with opposite spins, $a^\dagger_{\bf m\sigma}a_{\bf
m',-\sigma}$, which also give zero on averaging and therefore were
omitted. Terms which are proportional to $J^2$ and describe multiple
spin-excitation scattering processes were not included into $A_2$
either -- in this article only the decay of the spin excitation into
the fermion pair is considered. This process is described by the
following terms:
\begin{eqnarray*}
A_2&=&\frac{4tJ}{N}\sum_{{\bf k}_1{\bf k}_2\sigma}g_{{\bf kk}_1{\bf
k}_2}\overline{a^\dagger_{{\bf k}_1\sigma}a_{{\bf k+k}_1-{\bf
k}_2,\sigma}s^z_{{\bf k}_2}},\\
g_{{\bf kk}_1{\bf k}_2}&=&\left(\gamma_{{\bf k}_2}+\frac{1}{4}\right)\\
&&\times(\gamma_{{\bf k}_2-{\bf k}_1}-\gamma_{{\bf k}_1}-\gamma_{{\bf
k+k}_1-{\bf k}_2}+\gamma_{{\bf k+ k}_1}),
\end{eqnarray*}
where the line over the operators indicates that in calculating
thermodynamic averages with $A_2$ by factorization, terms containing
couplings of hole operators from the same $A_2$ have to be omitted,
since such processes have already been included into $\omega^2_{\bf
k}s^z_{\bf k}$. Substituting $A_2$ into the above definition of
$\Pi({\bf k}\omega)$, neglecting the difference between $A_{2t}$ and
$A_2(t)$ and using the decoupling we get
\begin{eqnarray}
{\rm Im}\,\Pi({\bf k}\omega)&=&\frac{8\pi t^2J^2}{N^2(1-\gamma_{\bf
k})(J|C_1|+tF_1)}\frac{1-\exp(\omega/T)}{\omega}\nonumber\\
&\times&\sum_{{\bf k}_1{\bf k}_2}g^2_{{\bf kk}_1{\bf k}_2}
\int\!\!\!\!\int^\infty_{-\infty}d\omega_1d\omega_2n_B(\omega_2)
\nonumber\\
&\times&[1-n_F(\omega_1)]n_F(\omega+\omega_1-\omega_2)B({\bf
k}_2\omega_2)\nonumber\\
&\times&\Big[A({\bf k}_1\omega_1)A({\bf k+k}_1-{\bf
k}_2,\omega+\omega_1-\omega_2) \nonumber\\
&-&L({\bf k}_1\omega_1\sigma)L({\bf k+k}_1-{\bf
k}_2,\omega+\omega_1-\omega_2,\sigma)\Big]. \nonumber\\
&&\label{po}
\end{eqnarray}
Equation~(\ref{frequency}) is supposed to give a good approximation for
the real part of the frequency of spin excitations and therefore only
the imaginary part of $\Pi({\bf k}\omega)$ will be considered below.
Notice that ${\rm Im}\,\Pi({\bf k}\omega)$ is negative, finite for
$\omega=0$ and even with respect to the change of the sign of $\omega$.

As seen from Eq.~(\ref{po}), ${\rm Im}\,\Pi({\bf k}\omega)$ is finite
for ${\bf k}\rightarrow 0$, whereas $\omega_{\bf k}$ vanishes in this
limit. Therefore the spin Green's function~(\ref{sgf}) has a purely
imaginary, diffusive pole near the $\Gamma$ point, in compliance with
the result of the hydrodynamic theory. \cite{Forster} In the general
case properties of spin excitations near the $M$ point differ
essentially from those near $\Gamma$. In the calculations of
Ref.~\onlinecite{Sherman02} for the former excitations the real parts
of frequencies were larger than their imaginary parts due to the spin
gap. However, it is worth noting that in this comparison only the decay
into two fermions was considered as the source of damping. Another
source of damping -- multiple spin-excitation scattering -- was
neglected. However, even in the case of overdamped excitations with
${\bf k \approx Q}$ their frequencies will have real components due to
the spin gap.

To simplify further calculations we take into account that in the
considered underdoped case the spin spectral function $B({\bf
k}\omega)$ is strongly peaked near ${\bf Q}$ for $\omega \approx
\omega_{\bf Q}$. Allowing for the small value of $\omega_{\bf Q}$,
$A({\bf k+Q},\omega) \approx A({\bf k}\omega)$ and Eq.~(\ref{zsm}) we
get
\begin{eqnarray}
{\rm Im}\,\Pi({\bf k}\omega)&=&\frac{9\pi t^2J^2(1-x)}{2N(1-\gamma_{\bf
k})(J|C_1|+tF_1)}\nonumber\\
&&\times\sum_{\bf k'}(\gamma_{\bf k+k'}-\gamma_{\bf k'})^2
\nonumber\\
&&\times\int^\infty_{-\infty}d\omega'
\frac{n_F(\omega+\omega')-n_F(\omega')}{\omega}\nonumber\\
&&\times\Big[A({\bf k'}\omega')A({\bf k+k'},\omega+\omega')\nonumber\\
&&-L({\bf k'}\omega'\sigma)L({\bf k+k'},\omega+\omega',\sigma)\Big].
\label{apo}
\end{eqnarray}
Now the damping has taken the familiar form given by the fermion
bubble.

\section{Magnetic susceptibility}
We have used hole self-energies~(\ref{se}) and correlations of hole and
spin operators obtained in Ref.~\onlinecite{Sherman02} for calculating
the hole Green's function ${\bf G}$, Eq.~(\ref{hgf}). This function and
the spin gap parameters $\Delta$ obtained in
Ref.~\onlinecite{Sherman02} have then been applied for the calculation
of the spin Green's function determined by Eqs.~(\ref{sgf}),
(\ref{frequency}) and~(\ref{apo}). This latter function is connected
with the magnetic susceptibility by the relation
$$\chi^z({\bf k}\omega)=-4\mu_B^2 D({\bf k}\omega),$$ where $\mu_B$ is the
Bohr magneton. The self-energies of Ref.~\onlinecite{Sherman02} were
calculated for a 20$\times$20 lattice with the parameters $t=0.5$~eV,
$J=0.1$~eV which correspond to hole-doped cuprates
\cite{Jefferson,McMahan} and for the ranges of hole concentrations and
temperatures $0\leq x\leq 0.16$ and $0.01t\approx 58\,{\rm K}\leq T\leq
0.2t\approx 1200\,{\rm K}$. For several hole concentrations we have
checked now that the self-energies calculated for $T=0.01t$ remain
practically unchanged as the temperature decreases to $T=0.003t\approx
17\,{\rm K}$. Therefore we can use these self-energies also for
$T<0.01t$. For temperatures close to zero the superconducting gap
$\Delta^s$ was set to $0.04t=20$~meV, the value extracted from the
tunnelling experiments. \cite{Yeh} As follows from the experiments,
this value remains practically unchanged with the doping variation from
heavily underdoped to optimally doped YBa$_2$Cu$_3$O$_{7-y}$.

Results of such calculations for the imaginary part of the magnetic
susceptibility at the antiferromagnetic wave vector ${\rm Im}\chi({\bf
Q})$ are shown in Figs.~\ref{Fig_i} and~\ref{Fig_ii}. In these figures
experimental data \cite{Bourges} on the magnetic susceptibility of
underdoped YBa$_2$Cu$_3$O$_{7-y}$ are also depicted. The oxygen
deficiencies $y=0.5$ and $0.17$ in this crystal correspond to the hole
concentrations $x \approx 0.075$ and $0.14$. \cite{Tallon}
YBa$_2$Cu$_3$O$_{7-y}$ is a bilayer crystal and the symmetry allows one
to divide the susceptibility into odd and even parts. For the
antiferromagnetic intrabilayer coupling the odd part can be compared
with our calculations carried out for a single layer.
\begin{figure}
\vspace*{8cm}
%\centerline{\includegraphics[width=7.2cm]{Fig1.eps}}
\caption{\label{Fig_i}The imaginary part of the spin susceptibility at
the antiferromagnetic wave vector in the superconducting state. Curves
show the results of our calculations in a 20$\times$20 lattice for
$t=0.5$~eV, $J=0.1$~eV, $T=17$~K, $x=0.06$ (a) and $x=0.12$ (b). Filled
squares are the odd susceptibility measured \protect\cite{Bourges} in
YBa$_2$Cu$_3$O$_{6.5}$ (a, $T_c=45$~K, $x \approx 0.075$) and in
YBa$_2$Cu$_3$O$_{6.83}$ (b, $T_c=85$~K, $x \approx 0.14$) at $T=5$~K.
Here and in Fig.~\protect\ref{Fig_ii} tick labels on the vertical axes
correspond to the curves. In both figures experimental values are
approximately 1.5 times smaller than the calculated ones.}
\end{figure}
\begin{figure}
%\centerline{\includegraphics[width=7.2cm]{Fig2.eps}}
\vspace*{8cm}
\caption{\label{Fig_ii}The imaginary part of the spin
susceptibility in the normal state. Curves show the results of our
calculations for $T=116$~K, all other parameters are the same as for
the respective panels in Fig.~\protect\ref{Fig_i}. Filled squares are
the odd susceptibility measured \protect\cite{Bourges} in
YBa$_2$Cu$_3$O$_{6.5}$ (a) and in YBa$_2$Cu$_3$O$_{6.83}$ (b) at
$T=100$~K.}
\end{figure}

The value of damping $|{\rm Im}\Pi({\bf Q}\omega)|$ depends on widths
of peaks in the hole spectral functions near the Fermi surface. These
widths are determined by an artificial broadening which was introduced
in Ref.~\onlinecite{Sherman02} to stabilize the iteration procedure.
From the comparison with photoemission spectra \cite{Damascelli} of
YBa$_2$Cu$_3$O$_{7-y}$ it is seen that the peaks in
Ref.~\onlinecite{Sherman02} are more intensive and narrower than in
experiment which leads to a larger value and stronger frequency
dependence of the calculated damping. To weaken this difference and to
obtain a better fit of the shapes of the calculated susceptibility to
the experimental data we have decreased $|{\rm Im}\Pi({\bf Q}\omega)|$
by a factor $f$ and added a constant damping $\eta$ to it. This allows
us to weaken somewhat the frequency dependence of the total damping
$\Gamma({\bf Q}\omega)=|{\rm Im}\Pi({\bf Q}\omega)|/f+\eta$. As will be
discussed in greater details later, the low-frequency shoulder in ${\rm
Im}\chi({\bf Q}\omega)$ is connected with this dependence. Thus, the
fitting parameters $f$ and $\eta$ allow us to change the relative
intensity of this shoulder. The damping $\eta$ can be connected with
the processes of multiple spin-excitation scattering or scattering at
impurities. The frequency dependencies of the total damping used in the
calculation of the curves in Figs.~\ref{Fig_i} and~\ref{Fig_ii} are
shown in Fig.~\ref{Fig_iii}. Notice that the fitting parameters $f$ and
$\eta$ with the values given in the caption to this figure influence
only weakly the position of the maximum in susceptibility which is
determined by the value of $\omega_{\bf Q}$.
\begin{figure}
\vspace*{6.5cm}
%\centerline{\includegraphics[width=7.8cm]{Fig3.eps}}
\caption{\label{Fig_iii}The frequency dependence of the total damping
$\Gamma({\bf Q})$ used in the calculation of the four curves in
Figs.~\protect\ref{Fig_i} and~\protect\ref{Fig_ii}. The parameter $f$
is equal to 2.7 for the dashed curve and 2 for the other curves. The
parameter $\eta$ is equal to $0.027t$, $0.04t$, $0.012t$, and $0.029t$
for the solid, dashed, short-dashed and dash-dotted curves,
respectively.}
\end{figure}

As seen from Figs.~\ref{Fig_i} and~\ref{Fig_ii}, the position of this
maximum, the resonance peak, and its evolution with doping and
temperature described by the $t$-$J$ model are in good agreement with
those observed in YBa$_2$Cu$_3$O$_{7-y}$. In the model the maximum is
connected with the excitation of localized Cu spins at the
antiferromagnetic wave vector ${\bf Q}$. Its frequency $\omega_{\bf Q}$
determines the size of the spin gap. In the underdoped case it
determines also the frequency of the resonance peak. As shown in
Ref.~\onlinecite{Sherman02}, $\omega_{\bf Q}$ grows with doping and
this leads to the growth of the frequency of the resonance peak from
approximately 18~meV at $x=0.06$ to 38~meV at $x=0.12$ in
Figs.~\ref{Fig_i} and~\ref{Fig_ii}. It was also shown \cite{Sherman02}
that ${\rm Im}\chi({\bf k}\omega)$ is strongly peaked at ${\bf Q}$ and
that the value of ${\rm Im}\chi({\bf Q}\omega)$ decreases with doping
which is in agreement with experimental observations.
\cite{Rossat,Bourges} In absolute units our calculated values of ${\rm
Im}\chi({\bf Q}\omega)$ are approximately 1.5 times larger than its
experimental values.

We notice that the shape of the calculated frequency dependence of the
susceptibility is close to that observed experimentally. Of special
interest is the low-frequency shoulder in this dependence. This
shoulder is more pronounced for lower hole concentrations and
temperatures. As mentioned above, it originates from the strong
frequency dependence of the damping $\Gamma({\bf Q})$ shown in
Fig.~\ref{Fig_iii}. The pronounced maxima of the curves in this figure
are connected with intensive peaks in the hole spectral function for
momenta near the Fermi surface. These peaks correspond to the so-called
spin-polaron band. \cite{Izyumov} For moderate doping the Fermi surface
of the $t$-$J$ model consists of two rhombuses with rounded corners.
\cite{Sherman02} These rhombuses are centered at the $\Gamma$ and $M$
points and are approximately nested by the momentum ${\bf Q}$. This
nesting is also very essential for the appearance of the maximum in
$\Gamma({\bf Q}\omega)$. In Fig.~\ref{Fig_iv} the Fermi surface is
shown and momenta of the hole spectral functions which give the main
contribution to the maxima of $\Gamma({\bf Q})$ in the used
20$\times$20 lattice are indicated. For these momenta the intensive
spin-polaron maxima in the spectral functions $A({\bf k'}\omega')$ and
$A({\bf Q+k'},\omega+\omega')$ [see Eq.~(\ref{apo})] overlap and fall
into the frequency window determined by the difference of the
occupation numbers.
\begin{figure}
\vspace*{6.5cm}
%\centerline{\includegraphics[width=6cm]{Fig4.eps}}
\caption{\label{Fig_iv}The Fermi surface of the $t$-$J$ model (lines)
and the momenta which give the main contribution to the maxima of
$\Gamma({\bf Q})$ in the used 20$\times$20 lattice (circles). The
antiferromagnetic wave vector ${\bf Q}$ connecting momenta of the
fermion pair in the spin polarization bubble is shown by the arrow. The
point $M$ corresponds to ${\bf k}=(\pi,\pi)$.}
\end{figure}

The Fermi surface in YBa$_2$Cu$_3$O$_{7-y}$ differs from that shown in
Fig.~\ref{Fig_iv}. \cite{Damascelli,Andersen} However, it is known from
the photoemission experiments that at least in the superconducting
state the hole spectral function has pronounced peaks for momenta near
the Fermi surface. In the two-layer YBa$_2$Cu$_3$O$_{7-y}$ the main
contribution to the damping of the spin excitations is given by the
decay into the fermion pair in which one of the fermions belongs to the
bonding band and the other to the antibonding band and the respective
parts of the Fermi surface are nested by the momentum $(\pi,\pi,\pi)$.
\cite{Bulut,Andersen} These conditions are similar to those observed in
the $t$-$J$ model and therefore the low-frequency shoulder in the
susceptibility in YBa$_2$Cu$_3$O$_{7-y}$ can be also related to the
strong frequency dependence of the damping of the spin excitations
which arises due to pronounced peaks in the hole spectral function and
the nesting.

It is worth noting that for all four curves in Figs.~\ref{Fig_i}
and~\ref{Fig_ii} the value of $\Gamma({\bf Q},\omega_{\bf Q})/2$ is
smaller than $\omega_{\bf Q}$. Thus, in contrast to a vicinity of the
$\Gamma$ point near the $M$ point the spin excitations are not
overdamped in underdoped YBa$_2$Cu$_3$O$_{7-y}$.

Now let us consider the momentum dependence of the resonance mode. In
Fig.~\ref{Fig_v} the constant energy scans obtained in our calculations
are compared with experiment \cite{Bourges95} in
YBa$_2$Cu$_3$O$_{6.83}$. The scans were performed along the diagonal of
the Brillouin zone at the resonance energy in the superconducting and
normal states. To simulate a finite instrumental momentum resolution,
which is comparable to the width of the peak in ${\rm Im}\chi({\bf
k}\omega)$ our curves were calculated by the convolution of this
quantity with the Gaussian with the full width at half maximum equal to
$0.2\pi$ in the momentum space. This corresponds to 0.1 in reciprocal
lattice units which is the usual resolution in experiments of this
type. As can be seen from Fig.~\ref{Fig_v}, for both temperatures the
calculated momentum dependencies are in good agreement with experiment.
\begin{figure}
\vspace*{6.5cm}
%\centerline{\includegraphics[width=7.8cm]{Fig5.eps}}
\caption{\label{Fig_v} Constant energy $(\kappa,\kappa)$ scans at the
resonance energy $\omega_{\bf Q}$. Solid and dashed curves show the
results of our calculations for $x=0.12$, $\omega_{\bf Q}=38$~meV in
the superconducting state at $T=17$~K and in the normal state at
$T=116$~K, respectively. To simulate a finite instrumental momentum
resolution the curves were calculated by the convolution of ${\rm
Im}\chi({\bf k}\omega)$ with the Gaussian with the full width at half
maximum equal to $0.2\pi$ in the momentum space. Filled and open
squares are experimental data \protect\cite{Bourges95} in
YBa$_2$Cu$_3$O$_{6.83}$ for $\omega_{\bf Q}=35$~meV, $T=4$~K and 109~K,
respectively.}
\end{figure}

In Fig.~\ref{Fig_vi} the dispersion of the maximum of our calculated
susceptibility is compared with experimental data \cite{Bourges} in
YBa$_2$Cu$_3$O$_{6.5}$. This dispersion corresponds approximately to
$\omega_{\bf k}$ in Eq.~(\ref{frequency}). For small ${\bf q=k-Q}$ this
momentum dependence can be written as
\begin{equation}
\omega_{\bf k}\approx\sqrt{\omega^2_{\bf Q}+c^2({\bf k-Q})^2}.
\label{apdisp}
\end{equation}
This function fitted to our calculated data with the parameters
$\omega_{\bf Q}=18.4$~meV and $c/a=\sqrt{8\alpha|C_1|}J=0.134$~meV is
also shown in Fig.~\ref{Fig_vi}. Here $a$ is the distance between Cu
sites in a Cu-O plane. As seen from this figure, our calculated
dispersion is close to the experimental one for similar parameters. For
$\omega\approx\omega_{\bf Q}$ ${\rm Im}\chi({\bf k}\omega)$ is peaked
at ${\bf k=Q}$. For $\omega>\omega_{\bf Q}$ the susceptibility has
maxima on the ring with the radius approximately determined by the
equation $\omega=[\omega^2_{\bf Q}+c^2({\bf k-Q})^2]^{1/2}$. In
constant energy scans along some direction this property of the
susceptibility manifests itself as two peaks in incommensurate
positions equally spaced from the $M$ point. \cite{Fong} For
$\omega<\omega_{\bf Q}$ for the considered parameters ${\rm
Im}\chi({\bf k}\omega)$ is peaked at ${\bf k=Q}$.
\begin{figure}
\vspace*{6.5cm}
%\centerline{\includegraphics[width=7.8cm]{Fig6.eps}}
\caption{\label{Fig_vi} The dispersion of the maximum in the frequency
dependence of ${\rm Im}\chi({\bf q}\omega)$, ${\bf q=k-Q}$. Filled
squares are our results for $x=0.06$ and $T=17$~K. The fit for these
data with Eq.~(\protect\ref{apdisp}) is shown by the curve. Open
squares are experimental results \protect\cite{Bourges} in
YBa$_2$Cu$_3$O$_{6.5}$ at $T=5$~K for odd spin excitations.}
\end{figure}

\section{Concluding remarks}
We have considered the magnetic susceptibility for the underdoped case
when the resonance peak is observed both in the normal and in the
superconducting states. As mentioned, the frequency of the peak is
determined by the frequency of the spin excitation $\omega_{\bf Q}$
which sets the size of the spin gap. This frequency grows with the hole
concentration, \cite{Sherman02} in agreement with experimental
observations in underdoped crystals. \cite{Bourges,He}

For the normal-state $t$-$J$ model in the overdoped region it was shown
\cite{Sherman97} that the part of the magnon branch, which persisted at
lower doping at the periphery of the Brillouin zone, is suddenly
destroyed for $x \approx 0.17$ at $T=0$. This transition is accompanied
by the radical change of the hole spectrum: dispersion and distribution
of the spectral weight become close to the case of weakly correlated
fermions. This result corresponds to the sudden disappearance of the
resonance peak in the normal-state overdoped cuprates.
\cite{Bourges,He} One of the reasons for the transition in the $t$-$J$
model is the damping of the spin excitations which grows with doping. A
considerable decrease of the damping in the superconducting state can
restore the spin excitations near the $M$ point in the frequency range
$\omega \leq 2\Delta^s$. Such mechanism was considered in
Ref.~\onlinecite{Morr} where the magnetic susceptibility similar to
that given by Eqs.~(\ref{sgf}) and~(\ref{frequency}) was postulated and
the damping described by the fermion bubble of the type of
Eq.~(\ref{apo}) was used. Above the mentioned transition at $x \approx
0.17$ the hole spectrum of the $t$-$J$ model becomes similar to that
used in Ref.~\onlinecite{Morr} and the analogous outcome can be
expected here.

In contrast to the underdoped region, in the overdoped case the
frequency of the resonance peak decreases with doping which can be
related to a finite damping of the spin excitations and to the decrease
of the superconducting gap with doping in this range of concentrations.
\cite{Yeh}

In contrast to YBa$_2$Cu$_3$O$_{7-y}$ where $\omega_{\bf Q} \alt
2\Delta^s$, in La$_{2-x}$Sr$_x$CuO$_4$ the value of $2\Delta^s \approx
9$~meV is substantially smaller than $\omega_{\bf Q}$ which is supposed
to be approximately the same as in the former crystal. This difference
may be the reason for the absence of the resonance peak in overdoped
La$_{2-x}$Sr$_x$CuO$_4$. \cite{Morr} Changes in the susceptibility
observed \cite{Mason} in La$_{1.86}$Sr$_{0.14}$CuO$_4$ at the
superconducting transition consist of some suppression of ${\rm
Im}\chi$ below the superconducting gap and an increase above it. The
suppression can be connected with the decrease of the damping of the
spin excitation accompanying the opening of the gap, while the increase
of the signal above the gap is apparently a combined effect of the
transfer of the carrier spectral weight above the gap and the nesting
supposed \cite{Si} for the Fermi surface of this crystal.

In conclusion, we compared the magnetic susceptibility calculated in
the $t$-$J$ model with the experimental data in the underdoped
YBa$_2$Cu$_3$O$_{7-y}$. It was demonstrated that the calculations
reproduce correctly the frequency and momentum dependencies of the
experimental susceptibility and its variation with doping and
temperature in the normal and superconducting states. This allowed us
to interpret the maximum in the frequency dependence -- the resonance
peak -- as a manifestation of the excitation branch of localized Cu
spins and to relate the frequency of the maximum to the size of the
spin gap. The low-frequency shoulder well resolved in the
susceptibility of superconducting crystals was connected with a
pronounced maximum in the damping of the spin excitations. This maximum
is caused by intense quasiparticle peaks in the hole spectral function
for momenta near the Fermi surface and by the nesting.

\begin{acknowledgments}
This work was partially supported by the ESF grant No.~5548 and by DFG.
\end{acknowledgments}


\begin{thebibliography}{99}
\bibitem{Rossat}J.~Rossat-Mignot, L.~P.~Regnault, P.~Bourges,
P.~Burlet, C.~Vettier, and J.~Y.~Henry, in {\it Selected Topics in
Superconductivity}, edited by L.~C.~Gupta and M.~S.~Multani (World
Scientific, Singapore, 1993), p.~265.

\bibitem{Bourges}P.~Bourges, in {\it The Gap Symmetry and Fluctuations
in High Temperature Superconductors}, edited by J.~Bok, G.~Deutscher,
D.~Pavuna, and S.~A.~Wolf (Plenum Press, 1998), p.~349.

\bibitem{Regnault}J.~Rossat-Mignot, L.~P.~Regnault, C.~Vettier,
P.~Bourges, P.~Burlet, J.~Bossy, J.~Y.~Henry, and G.~Lapertot, Physica
C {\bf 185-189}, 86 (1991).

\bibitem{Mook}H.~A.~Mook, M.~Yethiraj, G.~Aeppli, T.~E.~Mason,
and T.~Armstrong, Phys.\ Rev.\ Lett.\ {\bf 70}, 3490 (1993).

\bibitem{Dai}H.~F.~Fong, B.~Keimer, D.~L.~Milius, and I.~A.~Aksay,
Phys.\ Rev.\ Lett.\ {\bf 78}, 713 (1997); P.~Dai, H.~A.~Mook, and
F.~Do\v{g}an, Phys.\ Rev.\ Lett.\ {\bf 80}, 1738 (1998).

\bibitem{He}H.~He, Y.~Sidis, P.~Bourges, G.~D.~Gu, A.~Ivanov,
N.~Koshizuka, B.~Liang, C.~T.~Lin, L.~P.~Regnault, E.~Schoenherr, and
B.~Keimer, Phys.\ Rev.\ Lett.\ {\bf 86}, 1610 (2001); P.~Bourges,
B.~Keimer, S.~Pailh\`es, L.~P.~Regnault, Y.~Sidis, and C.~Ulrich,
cond-mat/0211227 (unpublished)

\bibitem{Liu}D.~Z.~Liu, Y.~Zha, and K.~Levin, Phys.\ Rev.\ Lett.\
{\bf 75}, 4130 (1995).

\bibitem{Mazin}I.~I.~Mazin and V.~M.~Yakovenko, Phys.\ Rev.\ Lett.\
{\bf 75}, 4134 (1995).

\bibitem{Bulut}N.~Bulut and D.~J.~Scalapino, Phys.\ Rev.\ B {\bf 53},
5149 (1996).

\bibitem{Abanov}A.~Abanov and A.~V.~Chubukov, Phys.\ Rev.\ Lett.\ {\bf
83}, 1652 (1999).

\bibitem{Morr}D.~K.~Morr and D.~Pines, Phys.\ Rev.\ Lett.\ {\bf 81},
1086 (1998).

\bibitem{Birgeneau}M.~A.~Kastner, R.~J.~Birgeneau, G.~Shirane, and
Y.~Endoh, Rev.\ Mod.\ Phys.\ {\bf 70}, 897 (1998).

\bibitem{Jefferson}J.~H.~Jefferson, H.~Eskes, and L.~F.~Feiner, Phys.\
Rev.\ B {\bf 45}, 7959 (1992); A.~V.~Sherman, Phys.\ Rev.\ B {\bf 47},
11521 (1993).

\bibitem{Sherman02}A.~Sherman and M.~Schreiber, Phys.\ Rev.\ B {\bf 65},
134520 (2002); European Phys.\ J.\ B {\bf 32}, 203 (2003).

\bibitem{Izyumov}Yu.~A.~Izyumov, Usp.\ Fiz.\ Nauk {\bf 167}, 465
(1997) [Phys.-Usp.\ (Russia) {\bf 40}, 445 (1997)]; E.~Dagotto, Rev.\
Mod.\ Phys. {\bf 66}, 763 (1994).

\bibitem{Kondo}J.~Kondo and K.~Yamaji, Progr.\ Theor.\ Phys.\ {\bf 47},
807 (1972); H.~Shimahara and S.~Takada, J.\ Phys.\ Soc.\ Jpn.\ {\bf
60}, 2394 (1991).

\bibitem{Mermin}N.~D.~Mermin and H.~Wagner, Phys.\ Rev.\ Lett.\
{\bf 17}, 1133 (1966).

\bibitem{Sherman98}A.~Sherman and M.~Schreiber, in {\it Studies of High
Temperature Superconductors}, edited by A.~V.~Narlikar (Nova Science
Publishers, New York, 1999), vol.~27, p.~163; Physica C {\bf 303}, 257
(1998).

\bibitem{Forster}D.~Forster, {\it Hydrodynamic Fluctuations,
Broken Symmetry, and Correlation Functions} (W.~A.~Benjamin, Inc.,
London, 1975).

\bibitem{McMahan}A.~K.~McMahan, J.~F.~Annett, and R.~M.~Martin, Phys.\
Rev.\ B {\bf 42}, 6268 (1990); V.~A.~Gavrichkov, S.~G.~Ovchinnikov,
A.~A.~Borisov, and E.~G.~Goryachev, Zh.\ Eksp.\ Teor.\ Fiz.\ {\bf 118},
422 (2000) [JETP (Russia) {\bf 91}, 369 (2000)].

\bibitem{Yeh}N.-C.~Yeh, C.-T.~Chen, R.~P.~Vasquez, C.~U.~Jung,
S.-I.~Lee, K.~Yoshida, and S.~Tajima, J.\ Low Temp.\ Phys.\ {\bf 131},
435 (2003).

\bibitem{Tallon}J.~L.~Tallon, C.~Bernhard, H.~Shaked, R.~L.~Hitterman,
and J.~D.~Jorgensen, Phys.\ Rev.\ B {\bf 51}, 12911 (1995).

\bibitem{Damascelli}A.~Damascelli, Z.~Hussain, and Z.-X.~Shen,
Rev.\ Mod.\ Phys.\ {\bf 75}, 473 (2003).

\bibitem{Andersen}O.~K.~Andersen, O.~Jepsen, A.~I.~Liechtenstein, and
I.~I.~Mazin, Phys.\ Rev.\ B {\bf 49}, 4145 (1994).

\bibitem{Bourges95}P.~Bourges, L.~P.~Regnault, J.~Y.~Henry, C.~Vettier,
Y.~Sidis, and P.~Burlet, Physica B {\bf 215}, 30 (1995).

\bibitem{Fong}H.~F.~Fong, P.~Bourges, Y.~Sidis, L.~P.~Regnault,
J.~Bossy, A.~Ivanov, D.~L.~Milius, I.~A.~Aksay, and B.~Keimer, Phys.\
Rev.\ B {\bf 61}, 14773 (2000).

\bibitem{Sherman97}A.~Sherman, Phys.\ Rev.\ B {\bf 55}, 582 (1997).

\bibitem{Mason}T.~E.~Mason, A.~Schr\"oder, G.~Aeppli, H.~A.~Mook, and
S.~M.~Hayden, Phys.\ Rev.\ Lett.\ {\bf 77}, 1604 (1996).

\bibitem{Si}Q.~Si, Y.~Zha, K.~Levin, and J.~P.~Lu, Phys.\ Rev.\ B {\bf
47}, 9055 (1993); P.~B\'enard, L.~Chen, and A.-M.~S.~Tremblay, Phys.\
Rev.\ B {\bf 47}, 15217 (1993).
\end{thebibliography}
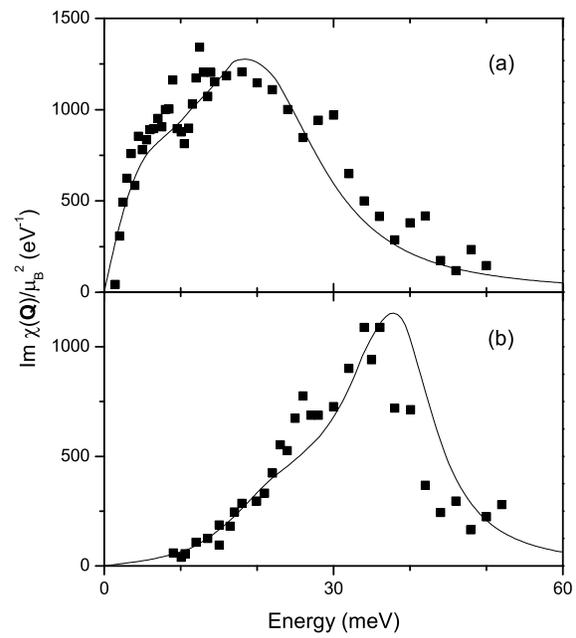
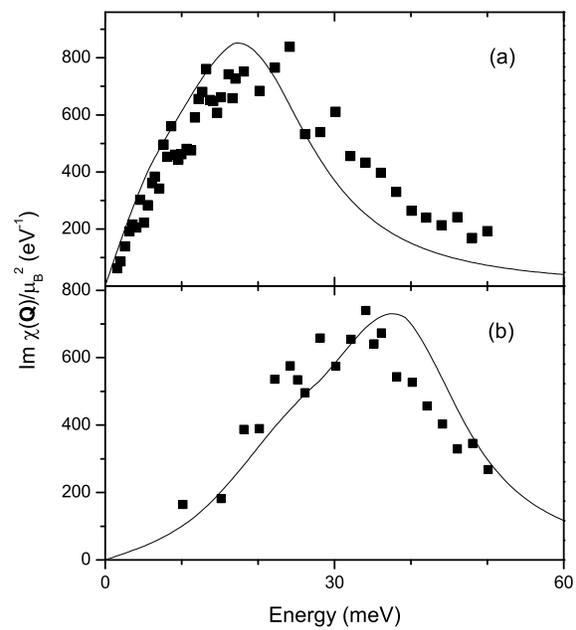
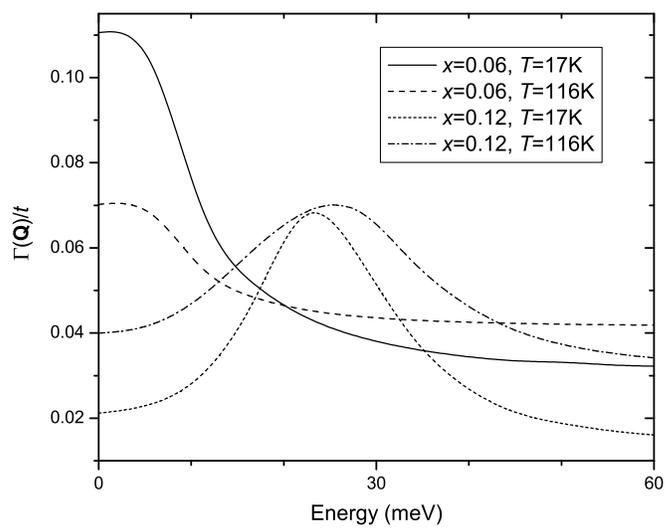
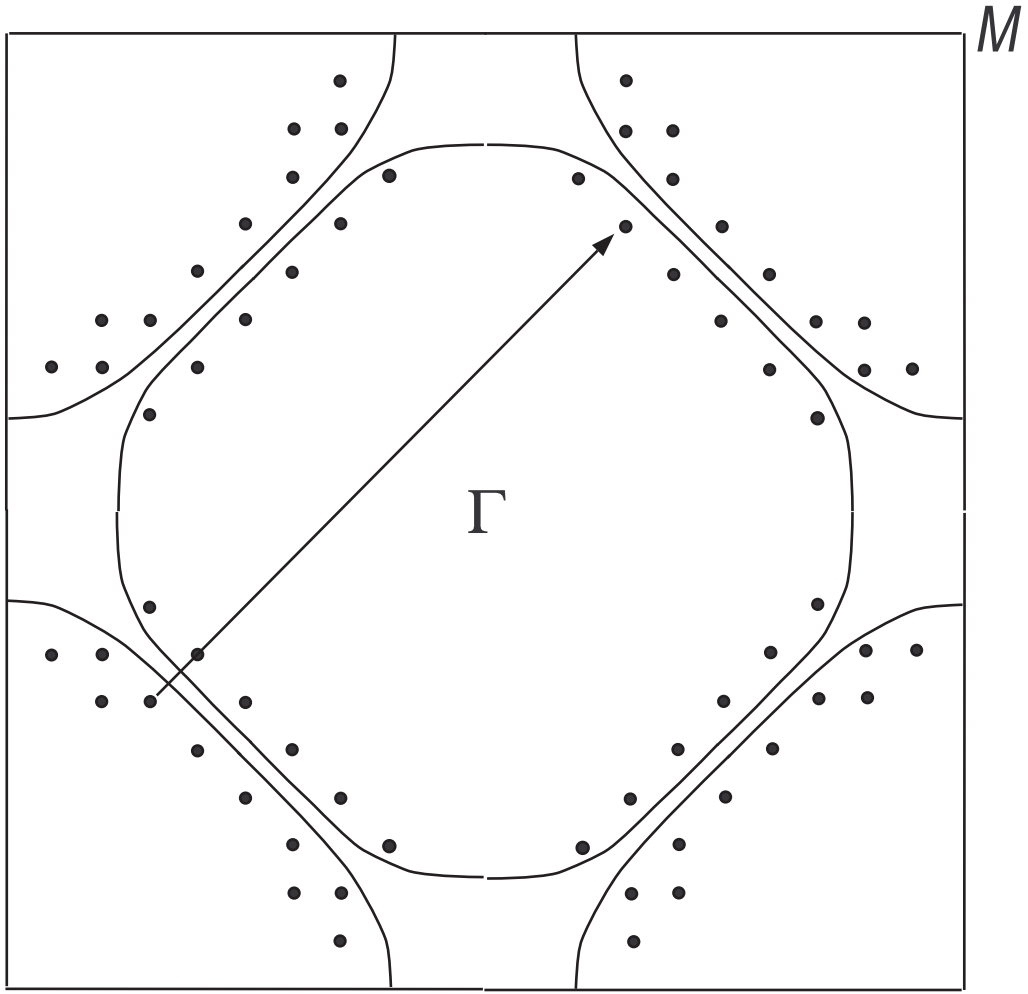
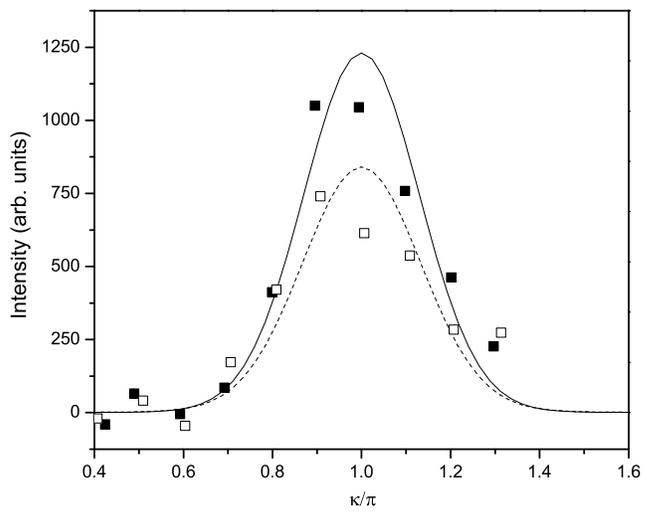
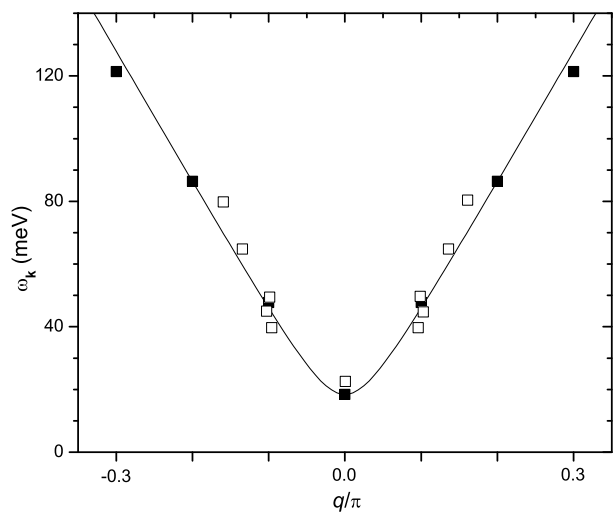
\end{document}